\documentclass[11pt]{article}
\usepackage{epsfig} \usepackage{amssymb} \usepackage{amsfonts}
\def\agoth{\relax\ifmmode{\mathfrak A}\else{${\mathfrak A}${ }}\fi}
\def\pisq{\relax\ifmmode{\pi^2}\else{${\pi^2}${ }}\fi}
\def\agothk{\relax\ifmmode{\mathfrak A}_k\else{${\mathfrak A}_k${ }}\fi}
\def\acal{\relax\ifmmode{\cal A}\else{${\cal A}${ }}\fi}
\def\acalk{\relax\ifmmode{\cal A}_k\else{${\cal A}_k${ }}\fi}
\newcommand{\beq}{\begin{equation}} \newcommand{\eeq}{\end{equation}}
\newcommand{\be}{\begin{equation}}
\newcommand{\ee}{\end{equation}}
\newcommand{\bea}{\begin{eqnarray}}
\newcommand{\eea}{\end{eqnarray}}
\begin{document}
\begin{center}

{\Large\bf  QCD coupling up to third order in standard and analytic perturbation theories}
\medskip

{\bf B.A. Magradze}
\footnote{\it Bogoliubov Laboratory of Theoretical Physics and A. Razmadze
 Tbilisi Mathematical Institute\\E-mail: magr@rmi.acnet.ge}

\end{center}

\section{Introduction}

In publications \cite{ss,ss0} a new approach to perturbative QCD, the
renormalization group (RG) invariant analytic approach (IAA) was
proposed (for the exhaustive review on this approach we recommend
paper \cite{ss1}). This method consistently takes into account the
renormalization invariance and analyticity in perturbation theory
(PT).

In work \cite{mss}, a particular version of IAA has been formulated.
 According this version, analytic perturbation
 theory (APT),
 the observables in the space-like region are represented as a
non-power series in the special universal
functions  ${\cal A}_{n}(Q^{2},f)$ (n=1,2\ldots and f denotes
number of quark flavor) \cite{sh1}. Analogical set of functions
\{$\agoth_n(s,f)$\}, $s>0$, was introduced in the time-like
region \cite{sh2}. Both sets of functions are determined  by the QCD running
coupling $\alpha_s(Q^2,f)$ and can be calculated in APT
analytically or numerically. A systematic mathematical investigation of these functions have been undertaken in works [5-7], in particular the oscillating
behavior in the infrared region was established.
To calculate these functions beyond the one-loop level the
iterative approximation for the coupling was used \cite{sh2,ms,mos}, or
RG equation for the coupling has been solved
numerically in the complex domain \cite{msy}.

In papers \cite{my1, ggk} the  RG equation for the QCD running
coupling, at the two-loop order, was solved explicitly as a
function of the scale. The solution was written in terms of
the Lambert W function. The three-loop order solution
(with Pade transformed $\beta$-function) also was expressed in
terms of the same function \cite{ggk}.
Using the explicit two-loop  solution, the analytical structure of the coupling
in the complex $Q^2$-plane has been determined \cite{my1,ggk}. The
analytical formulae for the corresponding spectral function was
found. Then the analytically improved coupling \cite{my2,my3} was reconstructed.

Afterwards, in paper \cite{cur}, the running coupling of an arbitrary
renormalization scheme, to the k-th order $(k\geq 3)$, was expanded as a power series in
the scheme independent explicit two-loop order solution. The new method for
reducing the scheme ambiguity for the QCD observables has been
proposed in this work. A similar expansion for the single scale dependent
observable, motivated differently, has been suggested in \cite{mx}.

In Sec.2 and Sec.3  we use the explicit solutions for the
running coupling to calculate   the universal quantities ${\cal
A}_{n}(Q^{2},f)$ and $\agoth_n(s,f)$ beyond the one loop order.
The results for $\agoth_n(s,f)$, to second and third orders,  are presented in the analytical form.

In Sec.4,  the matching conditions for
crossing the quark flavor thresholds for  the $\overline {MS}$ scheme
running coupling  $\alpha_{s}(Q,f)$, to the three loops, are solved
analytically. By the way we construct the global (independent on f)
universal functions, ${\cal A}_{n}(Q^{2})$ and $\agoth_{n}(s)$ (both introduced in \cite{sh2}). These functions can be used in the whole momentum space.

In Sec.5 we present numerical estimations of the   explicit
solutions for the coupling. The cases of standard PT and
 of APT are separately considered. We compare numerically 
Pade and the iterative approximants for the three-loop coupling. The scope of validity for the iterative approximant is estimated.
We give numerical results for the ``analyticized couplings'' ${\cal
A}_{1}(Q^{2},f)$ and $\agoth_{1}(s,f)$ to second and third orders.
 The differences between these
quantities and $\alpha_s(Q^2,f)$, are
estimated. The numerical results for the global  functions ${\cal
A}_{n}(Q^{2})$ and $\agoth_{n}(s)$ ($n=1,2,3$), to second and third orders, (see Tables 7-12) are given.

\section{Exact solution for the two-loop coupling  in the spacelike region}

The running coupling  of  QCD satisfies the differential equation
\footnote{We use the notation $Q^2=-q^2$, $Q^2>0$ corresponds to
a spacelike momentum transfer.}
\be
\label{eff}
Q^2 \frac{\partial{\alpha_{s}(Q^2,f)}}{\partial{Q^2}}=
{\beta}^{f}(\alpha_{s}(Q^2,f))=-\sum_{N=0}^{\infty}
\beta_{N}^{f}\alpha_{s}^{N+2}(Q^2,f),
\ee
${\alpha}_{s}({\mu}^2)={\alpha}_{s}=\frac{g^2}{4\pi}$, $g$ is
the renormalized coupling constant,  $\mu$ is the
renormalization point, and $f$ denotes the number of quark flavors.
In the class of schemes where the beta-function is mass
independent $\beta_0^{(f)}$ and $\beta_1^{(f)}$ are universal and the result for $\beta_{2}^{f}$ is available in the modified $MS$ $(\overline {MS})$ scheme
\be
\begin{array}{lll}
\label{coef}
\beta_{0}^{f}=\displaystyle\frac{1}{4\pi}\left(11-\frac{2}{3}f\right),&
\beta_{1}^{f}=\displaystyle\frac{1}{(4\pi)^2}\left(102-\frac{38}{3}f\right),&
\beta_{2}^{f}=\displaystyle\frac{1}{(4\pi)^3}\left(\frac{2857}{2}-\frac{5033f}{18}+\frac{325f^2}{54}\right).
\end{array}
\ee
For convenience in what follows we shall omit index f in
the coefficients $\beta_k^f$. Exact two-loop solution to
Eq.~(\ref{eff}) is given by \cite{my1,ggk}
\be
\label{w2}
\alpha_{s}^{(2)}(Q^2,f)=-\frac{\beta_0}{\beta_1}\frac{1}{1+
W_{-1}(\zeta)}:\hspace{5mm}
\zeta=-\frac{1}{eb}\left(\frac{Q^2}{\Lambda^2}\right)^{-\frac{1}{b}},
\ee
$b=\beta_1/\beta_0^2$, $\Lambda\equiv\Lambda_{\overline {MS}}$ and $W(\zeta)$ denotes the Lambert W
function \cite{lamb}, the multivalued inverse of
$
\zeta=W(\zeta)\exp W(\zeta).
$
The branches of W are denoted $W_{k}(\zeta), k=0,\pm 1,\ldots .$
A detailed review of  properties and
applications of this special  function can be found in
\cite{lamb}.
The three-loop solution (with Pade transformed beta-function) for the coupling is \cite{ggk}
\be
\label{w3}
\alpha_{Pade}^{(3)}(Q^2,f)=-\frac{\beta_0}{\beta_1}\frac{1}{1-\beta_0\beta_2/{\beta_1}^2+ W_{-1}(\xi)}:
\hspace{5mm}
\xi=-\frac{1}{eb}\exp\left(\frac{\beta_0\beta_2}{{\beta_1}^2}\right)\left(\frac{Q^2}{\Lambda^2}\right)^{-\frac{1}{b}}.
\ee
Expressions (\ref{w2}) and (\ref{w3}) allows us to perform analytical
continuation in the complex $Q^2$ plane and to calculate discontinuity along the
negative $Q^2$ axis. \footnote{For details of analytical continuation we recommend papers
[11-14].} In this  way   we  construct the corresponding
analytically improved  expressions for the coupling.  The ``analyticized n-th
power'' of the coupling, obtained from the solution (\ref{w3}), can be written as
\be
\label{an}
A^{(3)}_n(Q^2,f)\equiv\{\alpha_{s}^{(3)n}(Q^2,f)\}_{an.}=
\frac{1}{\pi}\int_{0}^{\infty}\frac{\rho_n^{(3)}(\sigma,f)}{\sigma+Q^2}
d\sigma=
\frac{1}{\pi}\int_{-\infty}^{\infty}\frac{e^{t}}{(e^{t}+Q^2/{\Lambda}^2)}{\tilde
\rho_n^{(3)}(t,f)}dt,
\ee
for $0\leq f \leq 6$, the spectral function, $\rho_n(\sigma,f)\equiv {\tilde
\rho_(t,f)}=\Im\{\alpha_{s}(-\sigma-\imath 0)\}^{n}$, is given by
\be \label{ro}
\tilde\rho_n^{(3)}(t,f)=\left(\frac{\beta_0}{\beta_{1}}\right)^{n}\Im\left(-\frac{1}{1-\beta_2\beta_0/\beta_1^2+
W_{1}(Z(t))}\right)^n, \ee with
\be
Z(t)=\frac{1}{be}\exp(\beta_0\beta_2/\beta_1^2-t/b+\imath(1/b-1)\pi).
\ee
Taking the limit $\beta_2\rightarrow 0$ in (\ref{ro}) one can reproduce the corresponding formula for the two loop case.

\section{Continuation of the QCD running coupling to the timelike region}

The APT approach allows us to define the QCD coupling in the timelike region
in a correct manner \cite{ms,mos,js,o}. Here, instead of common power series for a timelike observables,
there appears asymptotic series over the set of oscillating functions \{$\agoth_{n}(s,f)$\} \cite{sh2,sh3}.
The functions $\agoth_{n}(s,f)$ are defined by the elegant formula \cite{ms1}
\be
\label{tr}
\agoth_{n}(s,f)
=\frac{1}{\pi}\int_{s}^{\infty}\frac{d\sigma}{\sigma}\rho_n(\sigma,f),
\ee
here the spectral function is $\rho_n(\sigma,f)=\Im\{\alpha_{s}(-\sigma-\imath 0)\}^{n}$.
In the one-loop order the  set $\{\agoth_{n}(s)\}$ was
studied in Ref.\cite{sh2}.  In this case, the first four functions
of the set are given by
\be
\begin{array}{ll}
{\agoth_{1}^{(1)}}(s,f)=\displaystyle\frac{1}{\beta_0}(0.5-\displaystyle\frac{1}{\pi}\arctan(\displaystyle\frac{\ln\bar
s }{\pi})),&
{\agoth_{2}^{(1)}}(s,f)=\displaystyle\frac{1}{\beta_0^2}\displaystyle\frac{1}{(\ln^2\bar
s+\pi^2)},\nonumber\\
 {\agoth_{3}^{(1)}}(s,f)=\displaystyle\frac{1}{\beta_0^3}\displaystyle\frac{\ln\bar
s }{(\ln^2{\bar s}+{\pi}^2)^2},& {\agoth_{4}^{(1)}}(s,f)= \displaystyle\frac{1}{\beta_0^4}\displaystyle\frac{\ln^2{\bar s}-{\pi}^2/3}{(\ln^2{\bar s}+{\pi}^2)^3},\\
 \end{array}
\ee
 here $\bar s=s/\Lambda^2$.
In
this section we will calculate  $\agoth_{n}(s,f)$  to  second and third
orders. Let us define the auxiliary functions
\be
\label{ox} R_n(s,f)=\frac{1}{\pi}\int_{\ln\bar s}^{\infty}dt{\tilde
a}^n(t,f).
 \ee
where $\Im {\tilde a}^n(t,f)={\tilde \rho}_{n}(t,f)\equiv\rho_{n}(\sigma,f)$ with $\sigma=\exp(t)$.
 Then $\agoth_n(s,f)=\Im R_n(s,f)$.
The expressions for $\tilde a$, at the two and three loop orders, can be read
from (\ref{ro}).
In the two loop case
\be
\label{2l}
\begin{array}{lll}\tilde
a^{(2)}(t,f)=-\displaystyle\frac{\beta_0}{\beta_1}\displaystyle\frac{1}{1+W_1(z(t))}:&
z(t)=\displaystyle\frac{e^{-1-t/b+i\phi}}{b},&
\phi=\pi\left(\frac{1}{b}-1\right). \\
 \end{array}
\ee
 Integral (\ref{ox}), with (\ref{2l}), can be
 rewritten as a contour integral in the complex z-plane
\be
\label{oxx}
R_n^{(2)}(s,f)=p_n\int_{z_\epsilon}^{z_{s}}
\frac{dz}{z}\frac{1}{(1+W_{1}(z))^n}, \ee here
\be
\label{zs}
p_n=\frac{(-1)^n}{\pi}\frac{\beta_0^{n-2}}{\beta_1^{n-1}},\hspace{3mm}
z_s=\frac{1}{eb}(\bar
s)^{(-\frac{1}{b})}e^{\imath\phi},\hspace{3mm}
z_{\epsilon}=\epsilon e^{\imath\phi},\hspace{0.5cm}\phi=\pi\left(\frac{1}{b}-1\right),
\ee
and the limit
$\epsilon\rightarrow 0$ is assumed. Let us introduce the new
integration variable in (\ref{oxx}) \be\omega=W(z),\hspace{0.5cm}
d\omega=\frac{W(z)}{1+W(z)}\frac{dz}{z},\ee then
\be
\label{o}
R_n^{(2)}(s,f)=p_n\int_{W_{1}(z_{\epsilon})}^{W_{1}(z_{s})}
\frac{1+\omega}{\omega(1+\omega)^n}d\omega.\ee For $n>2$,
from (\ref{o}) we find the relation
\be
\label{atn1}
{\agoth}_{n}^{(2)}(s,f)=-\frac{\beta_0}{\beta_1}{\agoth}_{n-1}^{(2)}(s,f)+\frac{p_n}{(n-2)}\Im(1+W_{1}(z_s))^{2-n}.
\ee
Eq.~(\ref{atn1}) can be rewritten as the recurrence relation for
 ${\agoth} _{n}(s)$
\be
\label{rc}
\frac{\partial\agoth_{n-2}^{(2)}(s,f)}{\partial\ln s}=-(n-2)(\beta_{0}\agoth_{n-1}^{(2)}(s,f)+\beta_{1}\agoth_{n}^{(2)}(s,f))
, \ee
 formula (\ref{rc}) gives the generalization
of the similar one-loop order relation obtained in paper \cite{sh2}.
From (\ref{rc}) with the help of Eq.~(\ref{tr}) we find analogical formula for the spectral function
\be
\label{rc1}
\frac{\partial\rho_{n-2}^{(2)}(\sigma,f)}{\partial\ln \sigma}=-(n-2)(\beta_{0}\rho_{n-1}^{(2)}(\sigma,f)+\beta_1\rho_{n}^{(2)}(\sigma,f)).
\ee
Let us multiply Eq.~(\ref{rc1}) by the factor $(\sigma+Q^2)^{-1}$ and take the integral over the region $0<\sigma<\infty$. Integrating by parts and taking into account the condition
$
\rho_{n-2}(\sigma)\sigma/(Q^2+\sigma)|_{\sigma=0}^{\infty}=0
$
we obtain
\be
\label{rc2}
\frac{\partial \acal_{n-2}^{(2)}(Q^2,f)}{\partial \ln Q^2}=-(n-2)(\beta_{0}\acal_{n-1}^{(2)}(Q^2,f)+\beta_1\acal_{n}^{(2)}(Q^2,f)).
\ee
Note that for $n=3$ Eqs.~(\ref{rc}) and (\ref{rc2})
are analogous to the basic Eq.~(\ref{eff}) with $\alpha_s^{n}$ replaced by
$\agoth_n$ and $\acal_n$ respectively. 
We remark, that Eqs.~(\ref{rc})-(\ref{rc2}) 
could also be derived on general grounds using the RG equation 
(\ref{eff}) together with the APT prescription. For this derivation there is no necessity in the explicit (exact) solutions of the RG equation. In higher orders, similar equations (which generalize Eq.~(\ref{eff}) for $\acal_1$ and $\agoth_1$) can also be obtained \cite{my4}.
In particular, to the k-th order, the equation holds
\be
\label{rec4}
\frac{\partial \acal_{n}^{(k)}(Q^2,f)}{\partial \ln Q^2}=
-n\sum_{N=0}^{k-1}\beta_{N}\acal_{n+N+1}^{(k)}(Q^2,f),
\ee
the derivation of this equation will be given elsewhere \cite{my4}.

In the two-loop case, it is sufficient, to calculate
$\agoth_1$ and $\agoth_2$. Note that  $R_1^{(2)}(s,f)$ is divergent
in the limit $\epsilon\rightarrow 0$, (see (\ref{oxx}) ).  Nevertheless,  it has finite imaginary
part \footnote{for the asymptotic behaviour of the W function see paper
\cite{lamb}}. By direct calculation we find
\be
\label{f1}
\tilde\alpha^{(2)}(s,f)=\agoth_{1}^{(2)}(s,f)=\frac{1}{\beta_{0}}-\frac{1}{\pi\beta_0}\Im\ln W_{1}(z_s),
\ee
\be
\label{f2}
\agoth_{2}^{(2)}(s,f)=\frac{1}{\pi\beta_1}\Im\ln\left(\frac{W_{1}(z_s)}{1+W_{1}(z_s)}\right),
\ee
with  $z_s$ given by (\ref{zs}).
 Using  the known asymptotic behavior of the W-function
\cite{lamb},  in the limit $s\rightarrow 0$, we verify \cite{ss,ss0}
\be
\label{asy}
\begin{array}{lllll}
\tilde\alpha^{(2)}(s,f)=\agoth_{1}^{(2)}(s,f)\rightarrow \frac{1}{\beta_0},& and& \agoth_{n}^{(2)}(s,f)\rightarrow 0 & for
&n>1.\\
\end{array}
\ee
  Analogically in  three loop case we find
\be
\label{frs}
\label{pd1}
\agoth_{1}^{(3)}(s,f)=\tilde\alpha^{(3)}(s,f)=-\frac{1}{\pi\beta_0}\left(\frac{1}{\eta}\Im\ln(W_{1}(Z_s))+
(1-\frac{1}{\eta})\Im\ln(\eta+W_1(Z_s))-\pi\right),
\ee

\be
\label{pd2}
\agoth_{2}^{(3)}(s,f)=\frac{1}{\pi\beta_1}\left(\frac{1}{{\eta}^2}\Im\ln\left(\frac{W_{1}(Z_{s})}{\eta+W_{1}(Z_s)}\right)
-(1-\frac{1}{\eta})\Im\left(\frac{1}{\eta+W_1(Z_s)}\right)\right),
\ee
\be
\agoth_{n}^{(3)}(s,f)=-\frac{\beta_0}{\eta \beta_1}\agoth_{n-1}^{(3)}(s,f)+\frac{p_n}{\eta (n-2)}\Im(\eta+W_{1}(Z_s))^{2-n}+\frac{p_{n}}{n-1}(\frac{1}{\eta}-1)\Im(\eta+W_{1}(Z_s))^{1-n}
\label{pdn}
\ee
where $n>2$, $\eta=1-\beta_0\beta_2/\beta_1^2$ and
$$Z_s=\frac{1}{b}(s/\Lambda^2)^{-1/b}\exp(-\eta+\imath(1/b-1)\pi).$$
Note that  the ``analyticized'' perturbative expansions for timelike observables
(which contain specific functions   $\agoth_n$)
may be rewritten as power series in traditional coupling
$\alpha_{s}(s)$ with modified  by   $\pi^2$-factors
coefficients \cite{sh2,sh3}. Previously, these  modified power series have been obtained in \cite{rad,kr}. Application of the series
can be found in
papers [23-29]. Thus,
  ``$\pi^2$-effects'' for various timelike quantities
have been estimated in  paper \cite{kat}, in particular, it was found that the $\pi^2$-factors give dominating contributions to the coefficients of $R(s)=\sigma_{tot}(e^{+}e^{-}\rightarrow$ hadrons)$/\sigma(e^{+}e^{-}\rightarrow \mu^{+}\mu^{-})$. On the other hand,
in recent paper \cite{sh3} various timelike events was analyzed in the f=5 region. Higher-order ``$\pi^2$-effects'' have been taken into account properly.
It was found that the extracted values for $\alpha_{s}(M_{z}^2)$ are
influenced  significantly by these effects.

\section{Matching procedure and construction of the global space-like and time-like couplings}

In MS-like renormalization schemes important issue is how to introduce
the matching conditions for the strong coupling constant at the heavy
quarks thresholds. In literature few different recipes are known
(see for example \cite{sh2,sh3,mar,mik,rd,cht,gi}).
Here, we  follow works \cite{sh2,sh3}. Let us impose the continuity relations
\be
\label{th}
\alpha_s(M^2_f,\Lambda_{f-1},f-1)=\alpha_s(M^2_f, \Lambda_{f},f).
\ee
Inserting in (\ref{th}) the three-loop solution (\ref{w3}) we solve equation (\ref{th})
for $\Lambda_{f}$
\be
\label{mth}
\Lambda_{f}=M_{f}\{-b_{f}F(z_{f-1})\exp(\eta_{f}+F(z_{f-1}))\}^{b_{f}/2}
\ee
where $b_f=\beta_1^f/(\beta_0^f)^2$, $\eta^f=1-\beta_0^f\beta_2^f/(\beta_1^f)^2$,
\be
z_{f-1}=-\frac{\exp(-\eta_{f-1})}{b_{f-1}}\left(\frac{\Lambda_{f-1}}{M_{f}}\right)^{2/b_{f-1}},
\ee
\be
F(z_{f-1})=(\eta_{f-1}+W_{-1}(z_{f-1}))\frac{\beta_1^{f-1}}{\beta_{0}^{f-1}}
\frac{\beta_0^{f}}{\beta_{1}^{f}}-\eta_{f}. \ee
 In paper \cite{sh2}
special model for the spectral functions $\rho_n(\sigma)$ was proposed
\be
\label{mdl}
\rho_n(\sigma)=\rho_n(\sigma,\Lambda_3,3)+\sum_{f\ge
4}\Theta(\sigma-M_f^2)(\rho_n(\sigma,\Lambda_f,f)-\rho_n(\sigma,\Lambda_{f-1}, f-1)),
\ee
here the mass $M_f$ corresponds to the quark with flavor f, and $\Lambda_f$
is determined according formula (\ref{mth}).
Inserting (\ref{mdl}) in formula  (\ref{tr})
  we find following expression for the ``analyticized powers'' of the global
coupling in the timelike region \cite{sh2}
 \bea \label{glt}
 \agoth_n(s)=\agoth_n(s,f)+C_n(f)& for &M_f\leq {\sqrt s} \leq M_{f+1},
\eea
where the shift coefficients $C_n(f)$
are defined by relation
\be
C_n(f)=\agoth_n(M_{f+1}^2,f+1)-\agoth_n(M_{f+1}^2,f)+C_k(f+1)
\ee
with $C_n(6)=0$.
 The analogical formula follows for the corresponding global spacelike functions ${\cal A}_n(Q^2)$
\begin{eqnarray}
\label{gls}
 {\cal A}_n(Q^2)=
\frac{1}{\pi}\int_0^{M_4^2}\frac{d\sigma}{\sigma+Q^2}\rho_n(\sigma,\Lambda_3,3)+\nonumber\\
\frac{1}{\pi}\sum_{f=4}^{5}\int_{M_f^2}^{M_{f+1}^2}\frac{d\sigma}{\sigma+Q^2}\rho_n(\sigma,\Lambda_f,f)+
\frac{1}{\pi}\int_{M_6^2}^{\infty}\frac{d\sigma}{\sigma+Q^2}\rho_n(\sigma,\Lambda_6,6)
\end{eqnarray}

\section{Numerical estimations}
For the quarks masses throughout  this paper we assume the values
$M_1=M_2=M_3=0$, $M_4=1.3$ $GeV$, $M_5=4.3$ $GeV$ and $M_6=170$ $GeV$.

 In practice, usually, the iterative approximation for
the coupling \cite{bt} is used
 \bea \label{it}
\alpha_{it}^{(3)}(Q^2,f)=\frac{1}{\beta_0
L}-\frac{\beta_1}{\beta_0^3}\frac{\ln L}{L^2}+\frac{1}{\beta_0^3
L^3}\left(\frac{\beta_1^2}{\beta_0^2}(\ln^2 L-\ln
L-1)+\frac{\beta_2}{\beta_0}\right), \eea
where $L=\ln Q^2/{\Lambda}^{2}_{\overline MS}$. The same formula is used in the timelike region. It is
instructive to compare solutions (\ref{w3}) and (\ref{it}) with exact numerical solution of the RG equation $\alpha_{nm}^{(3)}(Q^2,f)$.
Numerical results for these functions are summarized in the Table
2. We see that, for $Q>1$ $GeV$, the expression (\ref{w3}) is more accurate then (\ref{it}) and the difference between these functions becomes
noticeable for $Q<1.2$ $ GeV$.

In Table 3 we give results for the three-loop Pade approximated coupling $\alpha^{(3)}(Q^2,5)$ and the corresponding analytic coupling
 ${\cal A}_{1}^{(3)}(Q^2,5)$.  In Table 4,  $\alpha^{(3)}(s,5)$ and
${\agoth}_{1}^{(3)}(s,5)$ are compared.
  The interval 5 $GeV< Q$, $\sqrt s < 200$ GeV is chosen and it is assumed that
$\Lambda_5=264$ $MeV$. Here, we observe the inequality
\be
\label{inq}
{\alpha}^{(k)}(Q^2,5)>{\cal A}_{1}^{(k)}(Q^2,5)>{\agoth}_{1}^{(k)}(Q^2,5),
\ee
the relative difference $\Delta$ (\%) between ${\alpha}^{(3)}(Q^2,5)$ and  ${\cal A}_{1}^{(3)}(Q^2,5)$ decreases from 1.4\% at $Q=5$ $GeV$ to 0.15\% at $Q=200$ $GeV$, whereas the difference between
${\alpha}^{(3)}(s,5)$ and ${\agoth}_{1}^{(3)}(s,5) $ is more
appreciable: $\Delta(\%)=7.5\%$ at $\sqrt s=5$ $GeV$ and
$\Delta(\%)=1.6\%$ at $\sqrt s=200$ $GeV$.
The `` mirror symmetry '' \cite{ms}
between ${\cal A}_{1}^{(3)}(Q^2,5)$ and ${\agoth}_{1}^{(3)}(s,5) $ is essentially violated (see Tables 3 and 4). Thus, the relative difference,
$\Delta(\%)=({\cal A}_{1}^{(3)}(Q^2,5)-{\agoth}_{1}^{(3)}(s,5))/{\cal A}_{1}^{(3)}(Q^2,5)*100$, monotonically decreases from 5.6 \%  at $Q=\sqrt s=1$ $GeV$ to $1.5\%$ at 200 $GeV$.

In Table 5 various functions at the two-loop and three-loop orders are compared. We choose  $f=5$ and $\Lambda_5=215$ $MeV$. Then, from the matching formula (\ref{mth}), in the two-loop case $(\eta=1)$ we find
 $\Lambda_3=363$ $MeV$, while
$\Lambda_3=340$ $MeV$ in the three-loop case. The interval \\
$20$ $GeV<Q,{\sqrt s}<170$ $ GeV$ is chosen. Here, we demonstrate the
stability of  results of PT and of APT, with respect to the
higher-loop corrections. In this region, the numerical difference between
the three-loop and the corresponding two- loop couplings are of order
$0.3\%-0.2\%$.

In Tables 7-12 we have summarized numerical results for the global three-loop
 functions , ${\cal A}_{n}^{(3)}(Q^2)$ and
${\agoth}_{n}^{(3)}(s)$ for n=1,2,3. The values for the parameter $\Lambda_3$
are chosen at $350$ $MeV$, $400$ $MeV$, and
$450$ $MeV$.
The corresponding values for the $\Lambda_f$ for $f>3$ are calculated
 from the three-loop matching condition (\ref{mth}),  (see Table 1).

Let us compare the global three-loop function ${\cal
A}_{1}^{(3)}(Q^2)$ (see Table 9) and  ${\alpha}^{(3)}(Q^2,5)$
(Table 3) in the case $\Lambda_3=400$ $MeV$ (the corresponding value for
$\Lambda_5$ is 264 $ MeV$). We see that the global coupling ${\cal
A}_{1}^{(3)}(Q^2)$  does not obey the inequality (\ref{inq}). The
difference ${\alpha}^{(3)}(Q^2,5)-{\cal A}_{1}^{(3)}(Q^2)$ becomes
negative when $Q$ increases. For $\Lambda_3=400$ $MeV$, this takes place at $Q\sim 17$ $GeV$.
 This  difference is small but it   increases with Q. It is about 0.6\% at
$Q=200$ $GeV$. The same effect is occurred for other values of
the $\Lambda_3$.  This enhancement of the global APT coupling can be explained   from
formula (\ref{gls}). It is easy to verify that
(\ref{gls}) contains additional positive non-perturbative
contribution, which was not occurred in the case of the local APT coupling (\ref{an}).
This contribution increase the coupling for large values of Q.

  Such a behavior does not occur in the case of the
global timelike coupling ${\agoth}_{n}^{(3)}(s)$
(compare  Tables 4 and  10). However, the difference between
${\alpha}^{(3)}(s,f=5)$ and ${\agoth}_{1}^{(3)}(s)$ is more
appreciable, it is about 7\% at $s=5$  $GeV$,  3.4\% at $s=20$
$GeV$   and  1.75\%  at $s=90$ $GeV$. It is about 1\% at $s=200$ $GeV$.
The relative difference between ${\cal A}_{1}^{(3)}(Q^2)$ and ${\agoth}_{1}^{(3)}(s)$
decreases from 6.9\% at $Q=\sqrt s=2$ $GeV$ to 2\% at 100 $GeV$ (see Tables 9 and 10). The relative differences $\Delta_n$ (\%) between ${\cal A}_n^{(3)}(Q^2)$ and ${\agoth}_{n}^{(3)}(s)$, for n=2 and 3, are even large.
Thus, $\Delta_2$ (\%)=6.1\% and $\Delta_3$ (\%)=11.1\% at $Q=\sqrt s=100 $ $GeV$.

 With the algebraic computer system Maple V release 5 we were
able to calculate the quantities $\{\alpha_{s}(Q^2,f)\}^{n}$ and
$\agoth_{n}(s,f)$ (see formulas
(\ref{w2}),(\ref{w3}),(\ref{pd1})-(\ref{pdn}))  with any arbitrary
given accuracy. However, this is no case for formulas (\ref{an}) and (\ref{gls}).
These integrals are singular at $t\rightarrow\pm\infty$, therefore, one
needs to use a cutoff. With Maple V release 5 the cutoff may be as
big as $10^4$. This  guarantees  to obtain 4-5 reliable
digits after decimal point. Most of our calculations we have performed with this precision. To obtain more accurate results we suggest following formula
\be
\label{cut} {\cal A}_{n}(Q^2,f)={\cal
A}_{n}(Q^2,f,R)+{\agoth}_{n}(\Lambda^{2}e^{R},f)+O(e^{-R}), \ee
 Here
${\cal A}_{n}(Q^2,f,R)$ denotes the regulated integral (\ref{an}): the integral is taken over the finite region $-R\leq t \leq R$.  We remark that formula (\ref{cut})
provides sufficiently high precision  even for  low values of
the cutoff.
In addition, for practical calculations, here we suggest the formula
\be
\label{D}
{\cal A}_{n}(Q^2,f)=Q^2\int_{0}^{\infty}\frac{ds}{(s+Q^2)^2}{\agoth}_{n}(s,f),
\ee
evidently, this relation is also valid for the global quantities, $\acal_{n}(Q^2)$ and $\agoth_{n}(s)$.

\section{Conclusion}

The``analyticized powers'' of the coupling in the spacelike region $\acal_n^{(k)}(Q^2,f)$, to second and third orders, are written in terms
of the Lambert W function (see integrals (\ref{an}) and (\ref{D})).

The ``analyticized powers'' of the coupling in the timelike region,  ${\agoth}_{n}^{(k)}(s,f)$ (k=2,3),
are analytically calculated in terms of the Lambert W function
(see formulas (\ref{f1})-(\ref{pdn})). The recurrence relations
     for ${\agoth}_{n}^{(2)}(s,f)$ and
 $\acal_n^{(2)}(Q^2,f)$ are derived (see formulas (\ref{rc}) and (\ref{rc2})).

The matching conditions for crossing the quark flavor thresholds
(\ref{th}) are solved explicitly for $\Lambda_f$
in the cases of the exact 2-loop
and Pade improved 3-loop solutions (see formulas (\ref{w2}), (\ref{w3}) and (\ref{mth})).

The global model for the coupling, of Refs.\cite{sh2}-\cite{sh3}, is considered up to the third order in the context of obtained explicit solutions.

In Sec.5 numerical estimations of the explicit solutions for the powers of the standard and analytical couplings are given. We have compared various solutions
in the
large region of momentum transfer and energy, 1 $GeV$ $\leq{\sqrt s},Q\leq 200$ $GeV$
(see Tables 1-12).
We have confirmed  that the differences between the powers of the standard
iterative solution (\ref{it}) and the
``analyticized timelike powers''  ${\agoth}_{n}^{(3)}(s)$
are appreciable even for moderate energies.

.

\section{Acknowledgments}

The author would like to thank D.V. Shirkov for numerous discussions,
suggestions and useful criticism from which this paper developed.
It is pleasure to acknowledge collaboration with V.P.~ Gerdt,
 D.S. Kourashev, I.L. Solovtsov, O.P.~ Solovtsova and A.V. Sidorov.
I am grateful to A.L.~ Kataev and A.A. Pivovarov for critical comments and stimulating discussions.
Useful conversations with  A.M. Khvedelidze and N.V. Makhaldiani kindly acknowledged.
I would like to thank D.I.~ Kazakov and A.A. Vladimirov for their support during my stay in Dubna and
 the staff of the Bogoliubov Laboratory for their hospitality.

{\bf Note Added}\\
As I was writing this I was informed by   D.S.
Kourashev  that he   has also obtained the  analytical expressions for the
 ``analyticized powers'' of coupling in the timelike region   in the equivalent form \cite{kour1} (see formulas (\ref{f1})-(\ref{pdn})).

\begin{table}[h]
\caption{The values of $\Lambda_f$ determined by the threshold matching
condition (\ref{mth}). $\Lambda_3$  is chosen as a basis quantity.}
\vspace{0.2cm}
\begin{center}
\begin{tabular}{|l|l|l|l|l|l|l|l|}
\hline
\multicolumn{4}{|c|}{The three loop  case}&
\multicolumn{4}{c|}{Exact two loop case}\\
\hline
 $\Lambda_3$ & $\Lambda_4$ &$\Lambda_5$ &$\Lambda_6$ &
 $\Lambda_3$ &$\Lambda_4$ &$\Lambda_5$ &$\Lambda_6$ \\
 $MeV$ & $MeV$& $MeV$ & $MeV$ & $MeV$ & $MeV$ & $MeV$ & $MeV$ \\
\hline
 300  &  258.4 &  183.0 &  76.2    &             300 &   248.4 &  170.4  &   69.2     \\
310  &  268.2 &  190.8 &  79.7    &              310&   257.6 &  177.3  &    72.3    \\
320  &  278.0 &  198.6 &  83.4    &              320&   266.8 &   184.3 &    75.5    \\
330  &  287.9 &  206.5 &  87.0    &              330&   276.0 &   191.4 &     78.7   \\
340  &  297.8 &  214.4 &  90.8    &              340&   285.3 &    198.5&    81.9     \\
350  &  307.8 &  222.5 &  94.6    &              350&   294.6 &    205.7&       85.2  \\
360  &  318.0 &  230.6 &  98.4    &              360&   303.9 &    212.9&       88.5  \\
370  & 328.0  & 238.8  & 102.3     &             370 &  313.2  &  220.1  &     91.9    \\
380  & 338.1  & 247.1  & 106.2     &             380 &  322.6  &  227.4  &     95.2    \\
390  & 348.4  & 255.5  & 110.2     &             390 &  332.0  &  234.8  &     98.7     \\
400  & 358.7  & 263.9  & 114.3     &             400 &  341.4  &  242.2  &     102.1     \\
410  & 369.0  & 272.5  & 118.4     &             410 &  350.9  & 249.6   &  105.6    \\
420  & 379.4  & 281.1  & 122.6     &             420 &  360.4  &  257.1  &  109.1    \\
430  & 389.9  & 289.8  & 126.8     &             430 &  369.9  &  264.7  &   112.7   \\
440  & 400.5  & 298.6  & 131.1     &             440 &  379.5  &  272.2  &  116.2    \\
450  & 411.1  & 307.5  & 135.4     &             450 &  389.0  &  279.9  &   119.8    \\
460  & 421.8  & 316.5  & 139.8     &             460 &  398.6  &  287.5  &   123.5    \\
470  & 432.6  & 325.6  & 144.3     &             470 &  408.3  &   295.2 &   127.2    \\
480  & 443.4  & 334.8  & 148.8     &             480 &  417.9  &   303.0 &   130.9    \\
490  & 454.4  & 344.1  & 153.4     &             490 &  427.6  &   310.8 &   134.6    \\
500  & 465.4  & 353.5  & 158.1     &             500 &  437.3  &   318.6 &   138.4     \\
 \hline
\end{tabular}
\end{center}
\end{table}

\begin{table}[h]
\caption{The three loop Pade, ${\alpha}_{pd}^{(3)}$, and the iterative, ${\alpha}_{it}^{(3)}$, couplings versus the exact numerical solution of the RG equation, ${\alpha}_{nm}^{(3)}$.
We take $\Lambda_3=400$ $MeV$ and the values of $\Lambda_f$ ($f>3$) are given in Table 1. We denote $\Delta_{pd}(\%)=|\alpha_{nm}^{(3)}-\alpha_{pd}^{(3)}|/\alpha_{nm}^{(3)}*100$,
, $\Delta_{it}(\%)=|\alpha_{nm}^{(3)}-\alpha_{pd}^{(3)}|/\alpha_{nm}^{(3)}*100$ and $\Delta_{p,i}(\%)=|\alpha_{pd}^{(3)}-\alpha_{it}^{(3)}|/\alpha_{pd}^{(3)}*100$}
\vspace{0.2cm}
\begin{center}
\begin{tabular}{|l|l|l|l|l|l|l|}
\hline
$Q$ $ GeV $   &
$  {\alpha}^{(3)}_{nm}(Q^2)$ &
  ${\alpha}^{(3)}_{pd}(Q^2)$ &
${\alpha}^{(3)}_{it}(Q^2)$ &
$\Delta_{pd} $ &
$\Delta_{it} $ &
$\Delta_{p,i} $ \\
GeV&
f=3&
f=3&
f=3&(\%)
&(\%)
& (\%)\\
\hline
 .8  & .76491 & .90931   & .88340 & 18.9 & 15.5& 2.9     \\
 .9  & .63323 & .68022   & .69179 & 7.4  &  9.3& 1.7      \\
1.0  & .55414 & .57854   & .58784 & 4.4  &  6.1& 1.6      \\
1.1  & .50028 & .51548   & .52195 & 3.0  &  4.3& 1.3      \\
1.2  & .46075 & .47126   & .47589 & 2.3  &  3.3& 1.       \\
1.3  & .43025 & .43803   & .44153 & 1.8  & 2.6 & .8      \\
1.4  & .40587 & .41191   & .41469 & 1.5  &  2.2& .7       \\
1.5  & .38583 & .39069   & .39301 & 1.3  &  1.9& .6       \\
1.6  & .36901 & .37302   & .37503 & 1.1  &  -1.6& .5       \\
1.7  & .35464 & .35803   & .35982 & 1.   & 1.5 & .5      \\
1.8  & .34220 & .34512   & .34674 & .9   & 1.3 &  .5      \\
1.9  & .33130 & .33384   & .33535 & .8   & 1.2 &   .5     \\
2.0  & .32165 & .32389   & .32530 & .7   & 1.1 &  .4     \\
2.1  & .31303 & .31503   & .31637 & .6   & 1.1 & .4      \\
2.2  & .30527 & .30707   & .30835 & .6   & 1.0 &  .4     \\
2.3  & .29824 & .29988   & .30111 & .6   & 1.  & .4      \\
2.4  & .29184 & .29334   & .29452 & .5   & .9  &  .4      \\
2.5  & .28598 & .28735   & .28850 & .5   & .9  & .4       \\
2.6  & .28058 & .28185   & .28297 & .5   & .9   & .4     \\
\hline
$Q$ $ GeV $   &
$  {\alpha}^{(3)}_{nm}(Q^2)$ &
  ${\alpha}^{(3)}_{pd}(Q^2)$ &
${\alpha}^{(3)}_{it}(Q^2)$ &
$\Delta_{pd} $ &
$\Delta_{it} $ &
$\Delta_{p,i} $ \\
GeV&
f=4&
f=4&
f=4&(\%)
&(\%)
& (\%)\\
\hline
2  & .33260 & .33392   & .33392  & .4  & .4 &   0     \\
3  & .27479 & .27540   & .27565  & .2  & .3 &  .1     \\
4  & .24527 & .24565   & .24598  & .2  & .3 &  .1     \\
5  & .22664 & .22691   & .22726  & .1  & .3 &  .2     \\
6  & .21350 & .21372   & .21406  & .1  & .3 &  .2     \\
7  & .20359 & .20376   & .20410  & .1  & .3 &  .2     \\
8  & .19575 & .19590   & .19623  & .1  & .3 &  .2     \\
9  & .18935 & .18948   & .18980  & .1  & .2 &  .2     \\
10 & .18398 & .18410   & .18441  & .1  & .2 &  .2     \\
\hline
$Q$ $ GeV $   &
$  {\alpha}^{(3)}_{nm}(Q^2)$ &
  ${\alpha}^{(3)}_{pd}(Q^2)$ &
${\alpha}^{(3)}_{it}(Q^2)$ &
$\Delta_{pd} $ &
$\Delta_{it} $ &
$\Delta_{p,i} $ \\
GeV&
f=5&
f=5&
f=5&(\%)
&(\%)
& (\%)\\
\hline
10  & .18845 & .18849   & .18847  & .02 & .01&  .01     \\
15  & .17126 & .17128   & .17130  & .01 & .03&  .01     \\
20  & .16090 & .16092   & .16096  & .01 & .03&  .02     \\
25  & .15372 & .15373   & .15378  & .01 & .04&  .03     \\
30  & .14832 & .14833   & .14838  & .01 & .04&  .03      \\
\hline
\end{tabular}
\end{center}
\end{table}

\begin{table}[h]
\caption{The $Q^2$ dependence of the three loop Pade improved coupling
$ \alpha^{(3)}(Q^2,f)$, (\ref{w3}), and the corresponding analytic coupling ${\cal A}_{1}^{(3)}(Q^2,f)$ for $f=5$ and $\Lambda_5=264$ $MeV$ ($\Lambda_3=400$ $MeV$).
 $\Delta(\%)$ denotes the relative difference between the couplings.}
\vspace{0.2cm}
\begin{center}
\begin{tabular}{|l|l|l|l|l|l|l|l|}
\hline
$Q$ $ GeV $   &
$  {\alpha}^{(3)}(Q^2,5)$ &
$  {\cal A}_{1}^{(3)}(Q^2,5)$ &
$\Delta(\%) $ &
$ Q$ $ GeV $ &
$  \alpha^{(3)}(Q^2,5)$ &
$ {\cal A}_{1}^{(3)}(Q^2,5)$    &
$\Delta(\%) $ \\
\hline
5  &  .22814 &  .22494  &   1.4 &       60 &  .13094&  .13075 & .14 \\
6  & .21610  &   .21383 &   1.0 &       62 &  .13022&  .13003 & .14  \\
7  & .20692  &   .20521 &   .82 &       64 &  .12953&  .12934 & .14  \\
8  & .19959  &   .19825 &   .67 &       66 &  .12887&  .12868 & .14  \\
9  & .19357  &   .19247 &   .56 &       68 &  .12823&  .12805 & .14  \\
10 &  .18849 &  .18757  &   .48 &      70  & .12762 & .12744  &.14   \\
11 &  .18413 &  .18334  & .42   &       72 &  .12703&  .12686 & .14  \\
12 &  .18033 &  .17964  & .38   &       74 &  .12647&  .12629 & .14  \\
13 &  .17697 &  .17636  & .34   &       76 &  .12592&  .12575 & .14  \\
14 &  .17398 &  .17343  & .31   &       78 &  .12540&  .12522 & .14  \\
15 &  .17128 &  .17078  & .29   &       80 &  .12489&  .12471 & .14  \\
16 &  .16884 &  .16838  & .27   &       82 &  .12439&  .12422 & .14  \\
17 &  .16661 &  .16618  & .25   &       84 &  .12392&  .12374 & .14  \\
18 &  .16456 &  .16416  & .24   &       86 &  .12345&  .12328 & .14  \\
19 &  .16267 &  .16230  & .22   &       88 &  .12301&  .12283 & .14  \\
20 &  .16092 &  .16057  & .21   &       90 &  .12257&  .12240 & .14  \\
21 &  .15929 &  .15895  & .21   &       92 &  .12215&  .12198 & .14  \\
22 &  .15777 &  .15745  & .20   &       94 &  .12174&  .12157 & .14   \\
23 &  .15634 &  .15603  & .19   &       96 &  .12134&  .12117 & .14  \\
24 &  .15500 &  .15470  & .19   &       98 &  .12095&  .12078 & .14  \\
25 &  .15373 &  .15345  & .18   &      100 &  .1205 &  .1204  & .14   \\
26 &  .15254 &  .15226  & .18   &      105 &  .11967&  .11950 &  .14  \\
27 &  .15140 &  .15114  & .17   &      110 &  .11882&  .11865 &  .14  \\
28 &  .15033 &  .15007  & .17   &      115 &  .11803&  .11786 &  .14  \\
29 &  .14931 &  .14905  & .16   &      120 &  .11727&  .11710 &  .14  \\
30 &  .14833 &  .14808  & .16   &      125 &  .11656&  .11639 &  .14  \\
32 &  .14651 &   .14628 & .16   &      130 &  .11588&  .11571 &  .14  \\
34 &  .14485 &   .14462 & .15   &      135 &  .11523&  .11507 &  .14  \\
36 &  .14331 &   .14309 & .15   &      140 &  .11462&  .11445 &  .14  \\
38 &  .14189 &   .14167 & .15   &      145 &  .11403&  .11387 &  .14  \\
40 &  .14056 &   .14035 & .15   &      150 &  .11347&  .11331 &  .14  \\
42 &  .13933 &   .13912 & .14   &      155 &  .11294&  .11277 &  .14  \\
44 &  .13817 &   .13797 & .14   &      160 &  .11242&  .11225 &  .14  \\
46 &  .13709 &   .13689 & .14   &      165 &  .11193&  .11176 &  .14   \\
48 &  .13606 &   .13586 & .14   &      170 &  .11145&  .11128 &  .14  \\
50 &  .13509 &   .13490 & .14   &      175 &  .11099&  .11083 &  .14  \\
52 &  .13418 &   .13398 & .14   &      180 &  .11055&  .11038 &  .15   \\
54 &  .13331 &    .13312&  .14  &      185 &  .11012&  .10996 &  .15   \\
56 &  .13248 &    .13229&  .14  &      190 &  .10971&  .10955 &  .15   \\
58 &  .13169 &    .13150&  .14  &      195 &  .10932&  .10915 &  .15   \\
   &         &          &       &      200 &  .10893&  .10877 &  .15   \\
\hline
\end{tabular}
\end{center}
\end{table}

\begin{table}[h]
\caption{The three loop analytic coupling $\agoth_{1}^{(3)} (s,5)$
( see (\ref{pd1}))
versus the ordinary 3-loop Pade approximated
coupling $\alpha^{(3)}(s,5)$. We have assumed that $\Lambda_3=400$ $MeV$, correspondingly $\Lambda_5=264$ $MeV$.}
\vspace{0.2cm}
\begin{center}
\begin{tabular}{|l|l|l|l|l|l|l|l|}
\hline
${\sqrt s}$ $ GeV $   &
$  {\alpha}^{(3)}(s,5)$ &
  ${\agoth}_{1}^{(3)} (s,5)$ &
$\Delta(\%) $ &
$ \sqrt s $ $ GeV $ &
$  \alpha^{(3)}(s,5)$ &
$\agoth_{1}^{(3)} (s,5)$    &
$\Delta(\%) $ \\
\hline
 5  & .22814  & .21221  &  7.5   &      60 &  .13094 & .12793 &  2.3     \\
 6  & .21610  & .20253  &  6.7   &      62 &  .13022 & .12726 &  2.3      \\
 7  & .20692  & .19498  &  6.1   &      64 &  .12953 & .12662 &  2.2      \\
 8  & .19959  & .18887  &  5.6   &      66 &  .12887 & .12600 &  2.2      \\
 9  & .19357  & .18378  &  5.3   &      68 &  .12823 & .12541 &  2.2      \\
 10 &  .18849 &  .17945 &  5.0   &      70 &  .12762 & .12484 &  2.2      \\
 11 &  .18413 &  .17570 &  4.7   &      72 &  .12703 & .12429 &  2.2      \\
 12 &  .18033 &  .17241 &  4.5   &      74 &  .12647 & .12376 &  2.1      \\
 13 &  .17697 &  .16949 &  4.4   &      76 &  .12592 & .12325 &  2.1      \\
 14 &  .17398 &  .16687 &  4.2   &      78 &  .12540 & .12276 &  2.1      \\
 15 &  .17128 &  .16450 &  4.1   &      80 &  .12489 & .12228 &  2.1      \\
 16 &  .16884 &  .16234 &  4.0   &      82 &  .12439 & .12182 &  2.1      \\
 17 &  .16661 &  .16037 &  3.8   &      84 &  .12392 & .12137 &  2.0      \\
 18 &  .16456 &  .15855 &  3.7   &      86 &  .12345 & .12094 &  2.0      \\
 19 &  .16267 &  .15687 &  3.7   &      88 &  .12301 & .12052 &  2.0      \\
 20 &  .16092 &  .15530 &  3.6   &      90 &  .12257 & .12011 &  2.0      \\
 21 &  .15929 &  .15384 &  3.5   &      92 &  .12215 & .11972 &  2.0      \\
 22 &  .15777 &  .15247 &  3.4   &      94 &  .12174 & .11933 &  2.0       \\
 23 &  .15634 &  .15119 &  3.4   &      96 &  .12134 & .11896 &  2.0      \\
 24 &  .15500 &  .14998 &  3.3   &      98 &  .12095 & .11859 &  1.9      \\
 25 &  .15373 &  .14884 &  3.2   &     100 &  .12057 & .11823 &  1.9       \\
 26 &  .15254 &  .14776 &  3.2   &     105 &  .11967 & .11739 &  1.9       \\
 27 &  .15140 &  .14673 &  3.1   &      110&  .11882 & .11659 &  1.9      \\
 28 &  .15033 &  .14576 &  3.1   &      115&  .11803 & .11583 &  1.8       \\
 29 &  .14931 &  .14483 &  3.0   &      120&  .11727 & .11512 &  1.8       \\
 30 &  .14833 &  .14394 &  3.0   &      125&  .11656 & .11445 &  1.8       \\
 32 &  .14651 &  .14228 &  2.9   &     130 &  .11588 & .11381 &  1.8       \\
 34 &  .14485 &  .14076 &  2.9   &      135&  .11523 & .11320 &  1.8       \\
 36 &  .14331 &  .13935 &  2.8   &      140&  .11462 & .11261 &  1.7       \\
 38 &  .14189 &  .13805 &  2.7   &      145&  .11403 & .11206 &  1.7       \\
 40 &  .14056 &  .13683 &  2.7   &      150&  .11347 & .11153 &  1.7       \\
 42 &  .13933 &  .13570 &  2.6   &      155&  .11294 & .11102 &  1.7       \\
 44 &  .13817 &  .13463 &  2.6   &      160&  .11242 & .11053 &  1.7       \\
 46 &  .13709 &  .13363 &  2.5   &      165&  .11193 & .11006 &  1.6       \\
 48 &  .13606 &  .13268 &  2.5   &      170&  .11145 & .10961 &  1.6       \\
 50 &  .13509 &  .13179 &  2.5   &      175&  .11099 & .10917 &  1.6       \\
 52 &  .13418 &  .13094 &  2.4   &      180&  .11055 & .10875 &  1.6       \\
 54 &  .13331 &  .13013 &  2.4   &      185&  .11012 & .10835 &  1.6       \\
 56 &  .13248 &  .12936 &  2.4   &      190&  .10971 & .10796 &  1.6       \\
 58 &  .13169 &  .12863 &  2.3   &      195&  .10932 & .10758 &  1.6        \\
    &         &         &        &     200 &  10893  & .10721 &  1.6       \\
\hline
\end{tabular}
\end{center}
\end{table}

\begin{table}
\begin{center}
\caption{The two and three loop couplings ${\alpha}^{(k)}(Q^2,5)$,
 ${\cal A}_{1}^{(k)}(Q^2,5)$ and ${\agoth}_{1}^{(k)}(s,5)$ ($k=2,3$) as a
 functions of variables $Q$ and $\sqrt s$.
 We take f=5 and $\Lambda_{5}=215 $ MeV.} \vspace{0.2cm}
\begin{tabular}{|l|l|l|l|l|l|l|}\hline
Q,$\sqrt s$      & $ {\alpha}_{s}^{(2)}(Q^2,5) $ & ${\cal A}_{1}^{(2)}(Q^2,5)$   & ${\agoth}_{1}^{(2)} (s,5)$
      & $ {\alpha}_{s}^{(3)}(Q^2,5) $ & ${\cal A}_{1}^{(3)}(Q^2,5)$    &${\agoth}_{1}^{(3)}(s,5)$\\
GeV    &                               &                          &
    &                               &                          &    \\ \hline
20 &  .15373  &     .15345  &   .14888     & .15429     &  .15400     &     .14934       \\
25 &  .14719   &    .14696  &   .14294     & .14769     &  .14744      &    .14335        \\
30 &  .14226   &    .14205  &   .13843     & .14271     &  .14249      &    .13880         \\
35 &  .13835   &    .13815  &   .13483     & .13876     &  .13856      &    .13517         \\
40 &  .13514   &    .13495  &   .13185     & .13552     &  .13532      &    .13218          \\
45 &  .13243   &    .13224  &   .12934     & .13279     &  .13260      &    .12965         \\
50 &  .13010   &    .12992  &   .12717     & .13044     &  .13025      &    .12746         \\
55 &  .12806   &    .12788  &   .12527     & .12838     &  .12820      &    .12555         \\
60 &  .12626   &    .12608  &   .12358     & .12656     &  .12639      &    .12385         \\
65 &  .12464   &    .12447  &   .12207     & .12494     &  .12476      &    .12233         \\
70 &  .12318   &    .12301  &   .12070     & .12347     &  .12330      &    .12096         \\
75 &  .12186   &    .12169  &   .11946     & .12213     &  .12196      &    .11970             \\
80 &  .12064   &    .12047  &   .11831     & .12091     &  .12074      &    .11855           \\
85 &  .11952   &    .11936  &   .11726     & .11979     &  .11962      &    .11749         \\
90 &  .11849   &    .11832  &   .11628     & .11874     &  .11857      &    .11651           \\
95 &  .11753   &    .11736  &   .11538     & .11778     &  .11761      &    .11560           \\
100 &  .11663 &     .11646    &    .11453    & .11687     &  .11670      &  .11474           \\
105 &  .11579  &    .11562     &   .11373    & .11603     &  .11586      &  .11394             \\
110 &  .11500  &    .11483     &   .11298    & .11523     &  .11506      &  .11319             \\
115 &  .11425  &    .11408     &   .11228    & .11448     &  .11431      &  .11248               \\
120 &  .11355  &    .11338     &   .11161    & .11377     &  .11360      &  .11181                \\
125 &  .11288  &    .11271     &   .11098    & .11310     &  .11293      &  .11117                \\
130 &  .11225  &    .11208     &   .11037    & .11246     &  .11230      &  .11057                \\
135  & .11164  &    .11148     &   .10980    & .11186     &  .11169      &  .10999            \\
140  & .11107  &    .11090     &   .10926    & .11128     &  .11111      &  .10945              \\
145  & .11052  &    .11035     &   .10873    & .11073     &  .11056      &  .10892               \\
150  & .10999  &    .10983     &   .10823    & .11020     &  .11003      &  .10842             \\
155  & .10949  &    .10932     &   .10776    & .10969     &  .10953      &  .10794              \\
160  & .10901  &    .10884     &   .10730    & .10921     &  .10904      &  .10748                \\
165  & .10854  &    .10838     &   .10685    & .10874     &  .10857      &  .10703                  \\
170  & .10810  &    .10793     &   .10643    & .10829     &  .10813      &  .10660   \\ \hline
\end{tabular}
\end{center}
\end{table}

\begin{table}
\begin{center}
\caption{The shift constants $C_{k}(f)$ as a functions of the
parameter $\Lambda_3$.}
\vspace{0.2cm}
  \begin{tabular}{|l|l|l|l|l|l|}\hline
$\Lambda_3$ $MeV$& 250& 300& 350& 400& 450\\\hline
$C_{1}(3)$ &      .0110     &    .0137    &    .0169 &    .0203  &  .024 \\
$C_{2}(3)$ &     .0062      &   .0079     &     .0099&    .0120  &  .014 \\
$C_{3}(3)$ &      .0022     &     .0028   &    .0035 &     .0042 &  .0049 \\
$C_{1}(4)$ &      .0026     &      .0032  &    .0037 &    .0043  &  .0049 \\
$C_{2}(4)$ &        .0013   &        .0016&    .0019 &     .0023 &  .0027 \\
$C_{3}(4)$ &         .0004  &        .0005&    .0007 &     .0008 &  .0010 \\
$C_{1}(5)$ &         .0003  &        .0003&    .0003 &    .0083  &  .0004 \\
$C_{2}(5)$ &         .0001  &        .0001&    .0001 &      .0020&  .0001 \\
$C_{3}(5)$ &         .0000  &       .0000 &     .0000&     .0003 &  .0000 \\
\hline
\end{tabular}
\end{center}
\end{table}

\begin{table}[h]
\caption{The three-loop Pade approximated global  analytic
coupling ${\cal A}_{1}(Q^2)$ and the corresponding global
``analyticized  powers'' ${\cal A}_2(Q^2)$ and ${\cal A}_3(Q^2)$
 as a functions of the momentum transfer Q
for $\Lambda_{3}=350$ $MeV$. The matching conditions give
 $\Lambda_{4}=307.8$ $MeV$, $\Lambda_{5}=222.5$ $MeV$
and $\Lambda_{6}=94.6$ MeV.} \vspace{0.2cm}
\begin{center}
\begin{tabular}{|l|l|l|l|l|l|l|l|}
\hline $Q$ $GeV$&    ${\cal A}_{1}(Q^2)$&  $ {\cal A}_2(Q^2)$ &
   ${\cal A}_3(Q^2)$& $Q$ $GeV$ & ${\cal A}_{1}(Q^2)$&
    ${\cal A}_2(Q^2)$&   ${\cal A}_3(Q^2)$ \\
\hline
   1  &  .36571 &   .09320&   .015540 &        38 &    .13774   &    .01898  &  .00258 \\
   2  &  .28689 &   .06861&   .012776 &        40 &    .13650   &    .01864  &  .00252 \\
   3  &  .25144 &   .05641&   .010598 &        42 &    .13534   &    .01833  &  .00246 \\
   4  &  .23041 &   .04903&   .009107 &        44 &  .13442&     .018047 &  .00240 \\
   5  &  .21615 &   .04403&   .008042 &        46 &  .13340&      .01777 & .00235  \\
   6  &  .20566 &  .04039 &  .007245  &        48 &  .13244&      .01752 & .00230  \\
   7  &  .19753 &  .037601&  .006626  &        50 &  .13153&      .01728 & .00225  \\
   8  &  .19098 &  .035378&  .006131  &        52 &  .13067&      .01706 & .00221  \\
   9  & .18555  &  .033557&  .005724  &        54 &  .12986&      .01685 & .00217  \\
   10 & .18095  &  .032033&  .005384  &        56 &  .12908&      .01665 & .00213  \\
   11 & .17698  &  .030734&  .005095  &        58 &  .12834&      .01646 & .00210  \\
   12 & .17351  &  .029610&  .004845  &        60 &  .12763&      .01628 & .00206  \\
   13 & .17043  &  .028626&  .004628  &        62 &  .12696&      .01611 & .00203  \\
   14 & .16768  &  .027755&  .004436  &        64 &  .12631&      .01595 & .00200  \\
   15 & .16520  &  .026977&  .004266  &        66 &  .12569&      .01580 & .00197  \\
   16 & .16295  &  .026277&  .004113  &        68 &  .12510&      .01565 & .00195  \\
   17 & .16089  &  .025643&  .003975  &        70 &  .12452&      .01551 & .00192  \\
   18 & .15900  &  .050652&  .003851  &        72 &  .12397&      .01537 &  .00190 \\
   19 & .15725  &  .024535&  .003737  &        74 &  .12344&      .01524 &   .00187\\
   20 &  .15563  & .02405 &  .00363   &          76 &  .12293&      .01511 &   .00185 \\
   21 & .15412   & .02359 &  .00353   &         78  &  .12244&      .01499 &   .00183 \\
   22 & .15270   & .02317 &  .00344   &          80 &  .12196&      .01488 &   .00181 \\
   23 & .15138   & .02278 &  .00336   &          82 &  .12150&      .01477 &   .00179 \\
   24 & .15013   & .02242 &  .00328   &          84 &  .12105&      .01466 &   .00177 \\
   25 & .14895   & .02208 &  .00321   &          86 &  .12062&      .01455 &   .00175 \\
   26 & .14784   & .02176 &  .00315   &          88 &  .12020&      .01445 &   .00173 \\
   27 & .14678   & .02145 &  .00308   &          90 &  .11979&      .01436 &   .00171 \\
   28 & .14578   & .02117 &  .00303   &          92 &  .11940&      .01426 &   .00170 \\
   29 & .14483   & .02090 &  .00297   &          94 &  .11902&      .01417 &   .00168 \\
   30 & .14392   & .02064 &  .00292   &          96 &  .11864&      .01408 &   .00166 \\
   32 & .14222   & .02016 &  .00282   &          98 &  .11828&      .01400 &   .00165 \\
  34  & .14066   & .01973 &  .00273   &          100&  .11793&      .01391 &   .00163\\
  36  & .13923   & .01934 &  .00265   &             &        &             &         \\
 \hline
 \end{tabular}
 \end{center}
 \end{table}

\begin{table}[h]
\caption{The three-loop Pade approximated global  analytic coupling ${\agoth}_{1}(s)$ and
the corresponding  global ``analyticized  powers'' ${\agoth}_2(s)$ and ${\agoth}_3(s)$
       as a functions of the energy $\sqrt s $ for $\Lambda_{3}=350$
         $MeV$.}
\vspace{0.2cm}
\begin{center}
\begin{tabular}{|l|l|l|l|l|l|l|l|}
\hline ${\sqrt s}$ $GeV$ &   ${\agoth}_{1}(s)$ & $ {\agoth}_{2}(s)$ &
${\agoth}_3(s)$ & ${\sqrt s }$ $GeV$ &  ${\agoth }_{1}(s)$ & $ {\agoth}_{2}(s)$& ${\agoth}_3(s)$
\\
 \hline

1  &  .34270 &  .09196 & .01903 &     58 &  .12550 & .01544 &  .00185 \\
2  &  .26679 &  .06266 & .01286 &     60 &  .12484 & .01528 &  .00183 \\
3  &  .23451 &  .05024 & .00979 &     62 &  .12420 & .01513 &  .00180 \\
4  &  .21571 &  .04338 & .00811 &     64 &  .12359 & .01498 &  .00178  \\
5  &  .20343 &  .03896 & .00701 &     66 &  .12301 & .01484 &  .00175  \\
6  &  .19452 &  .03582 & .00623 &     68 &  .12245 & .01471 &  .00173  \\
7  &  .18755 &  .03344 & .00566 &     70 &  .12191 & .01459 &  .00171  \\
8  &  .18190 &  .03156 & .00521 &     72 &  .12139 & .01447 &  .00169  \\
9  &  .17718 &  .03002 & .00485 &     74 &  .12089 & .01435 &  .00167  \\
10 & .17316  & .02874  & .00456 &     76 &  .12040 & .01424 &  .00165  \\
11 & .16967  & .02764  & .00432 &     78 &  .11993 & .01413 &  .00163  \\
12 & .16661  & .02670  & .00411 &     80 &  .11948 & .01403 &  .00161  \\
13 & .16389  & .02587  & .00392 &     82 &  .11904 & .01393 &  .00160  \\
14 & .16144  & .02513  & .00376 &     84 &  .11862 & .01383 &  .00158  \\
15 & .15923  & .02447  & .00362 &     86 &  .11821 & .01373 &  .00156  \\
16 & .15722  & .02388  & .00350 &     88 &  .11781 & .01364 &  .00155  \\
17 & .15537  & .02335  & .00338 &     90 &  .11742 & .01356 &  .00153  \\
18 & .15367  & .02286  & .00328 &     92 &  .11704 & .01347 &  .00152   \\
19 & .15209  & .02241  & .00319 &     94 &  .11668 & .01339 &  .00151   \\
20 & .15063  & .02199  & .00311 &     96 &  .11632 & .01331 &  .00149   \\
21 & .14926  & .02161  & .00303 &     98 &  .11597 & .01323 &  .00148   \\
22 & .14798  & .02125  & .00296 &     100&  .1156  &.01316  & .00147    \\
23 & .14677  & .02092  & .00289 &     105&  .11483 & .01298 &  .00144   \\
24 & .14563  & .02061  & .00283 &     110&  .11407 & .01281 &  .00141   \\
25 &  .14456 &  .02031 &  .00277&     115 & .11335 &  .01265 &   .00139 \\
26 &  .14355 &  .02004 & .00271 &     120 & .11267 & .01251  & .00136   \\
27 &  .14258 &  .01978 & .00266 &     125 & .11203 & .01237  & .00134   \\
28 &  .14166 &  .01953 & .00262 &     130 & .11142 & .01224  & .00132   \\
29 &  .14079 &  .01930 & .00257 &     135 & .11084 & .01211  & .00130    \\
30 &  .13996 &  .01908 & .00253 &     140 & .11028 & .01199  & .00128    \\
32 &  .13840 &  .01867 &  .00245&     145 & .10975 & .01188  & .00126    \\
34 &  .13696 &  .01829 &  .00238&     150 & .10925 & .01177  & .00125     \\
36 &  .13564 &  .01795 &  .00231&     155 & .10876 & .01167  & .00123     \\
38 &  .13441 &  .01764 &  .00225&     160 & .10830 & .01157  & .00122     \\
40 &  .13326 &  .01735 &  .00220&     165 & .10785 & .01148  & .00120     \\
42 &  .13219 &  .01708 &  .00215&     170 & .10742 & .01139  & .00119     \\
44 &  .13118 &  .01682 &  .00210&     175 &  .10704&  .01131 &  .00118    \\
46 &  .13023 &  .01659 &  .00206&     180 &  .10667&  .01124 &  .00116    \\
48 &  .12933 &  .01637 &  .00202&     185 &  .10632&  .01116 &  .00115    \\
50 &  .12849 &  .01616 &  .00198&     190 &  .10598&  .01109 &  .00114    \\
52 &  .12768 &  .01596 &  .00195&     195 &  .10565&  .01102 &  .00113    \\
54 &  .12692 &  .01578 &  .00192&     200 &  .10533&  .01096 &  .00112    \\
\hline
\end{tabular}
\end{center}
\end{table}

\begin{table}[h]
\caption{The three-loop Pade approximated global  analytic coupling ${\cal A}_{1}(Q^2)$ and the corresponding ``analyticized  powers'' ${\cal A}_2(Q^2)$ and ${\cal A}_{3}(Q^2)$
 as a functions of the momentum transfer Q for $\Lambda_{3}=400$ $MeV$.
The matching conditions give
         $\Lambda_{4}=358.7$ $MeV$ and $\Lambda_{5}=263.9$ MeV.}
\vspace{0.2cm}
\begin{center}
\begin{tabular}{|l|l|l|l|l|l|l|l|}
\hline
$Q$ $GeV$&    ${\cal A}_{1}(Q^2)$&  $ {\cal A}_2(Q^2)$ &
   ${\cal A}_3(Q^2)$& $Q$ $GeV$ & ${\cal A}_{1}(Q^2)$&
    ${\cal A}_2(Q^2)$&   ${\cal A}_3(Q^2)$ \\
\hline
1  &   .38656 &   .09988 &  .01629 &     58 & .13190&   .01743 &    .00228 \\
2  &   .30263 &   .07438 &  .01387 &     60 & .13116&   .01723 &    .00224 \\
3  &   .26444 &   .06128 &  .01164 &     62 & .13044&   .01705 &    .00221  \\
4  &   .24171 &   .05325 &  .01005 &     64 & .12976&   .01687 &    .00217 \\
5  &   .22627 &   .04777 &  .00890 &     66 & .12910&   .01670 &    .00214  \\
6  &   .21492 &   .04377 &  .00803 &     68 & .12847&   .01654 &    .00211 \\
7  &   .20613 &   .04069 &  .00734 &    70 &  .12786 &   .01639 &    .00208   \\
8  &   .19905 &   .03824 &  .00679 &    72 &  .12728 &   .01624 &    .00206  \\
9  &   .19318 &   .03623 &  .00634 &    74 &  .12672 &   .01610 &    .00203  \\
10 &   .18822 &   .03454 &   .00596&    76 &  .12618 &   .01596 &    .00200  \\
11 &   .18394 &   .03311 &   .00564&    78 &  .12566 &   .01583 &    .00198  \\
12 &   .18020 &   .03187 &   .00536&    80 &  .12515 &   .01570 &    .00196  \\
13 &   .17689 &   .03078 &   .00490&    82 &  .12467 &   .01558 &    .00193  \\
14 &   .17393 &   .02982 &   .00471&    84 &  .12420 &   .01547 &    .00191  \\
15 &   .17127 &   .02896 &   .00454&    86 &  .12374 &   .01535 &    .00189  \\
16 &   .16885 &   .02819 &   .00438&    88 &  .12330 &   .01524 &    .00187  \\
17 &   .16664 &   .02749 &   .00424&    90 &  .12287 &   .01514 &    .00185  \\
18 &   .16461 &   .02685 &   .00411&    92 &  .12245 &   .01504  &    .00183  \\
19 &   .16273 &   .02627 &   .00400&    94 &  .12204 &   .01494  &    .00182  \\
20 &   .16099 &   .02573 &   .00389&    96 &  .12165 &   .01484  &    .00180  \\
21 &   .15937 &   .02523 &   .00379&    98 &  .12127 &   .01475  &    .00178  \\
22 &   .15786 &    .02477 & .00370 &    100 &  .12090 &   .01466 &     .00177 \\
23 &   .15644 &    .02434 & .00361 &    105 &  .12001 &   .01445 &     .00173 \\
24 &   .15510 &    .02394 & .00353 &    110 &  .11917 &   .01425 &      .00169 \\
25 &   .15385 &  .02357 &   .03460 &     115 &  .11839 &   .01406 &      .00166 \\
26 &   .15266 &  .02321 &   .00339 &     120 &  .11765 &   .01389 &      .00163 \\
27 &   .15153 &  .02288 &   .00332 &     125 &  .11695 &   .01372 &      .00160 \\
28 &   .15046 &  .02257 &   .00326 &     130 &  .11628 &   .01357 &      .00157 \\
29 &   .14944 &  .02227 &   .00320 &     135 &  .11565 &   .01342 &      .00155 \\
30 &   .14847 &  .02199 &   .00309 &     140 &  .11505 &   .01328 &      .00153 \\
32 &   .14666 &  .02147 &   .20099 &     145 &  .11448 &   .01315 &      .00150 \\
34 &   .14500 &  .02100 &   .00290 &     150 &  .11394 &   .01303 &      .00148 \\
36 &   .14347 &  .02056 &   .00282 &     155 &  .11341 &   .01291 &      .00146 \\
38 &   .14205 &  .02017 &   .00275 &     160 &  .11291 &   .01280 &      .00144 \\
40 &   .14074 &  .01980 &   .00269 &     165 &  .11243 &   .01269 &      .00142 \\
42 &   .13951 &  .01946 &   .00262 &     170 &  .11197 &   .01258&      .00141 \\
44 &   .13835 &  .01915 &   .00256 &     175 &  .11153 &   .01248 &      .00139 \\
46 &   .13727 &  .01885 &   .00251 &     180 &  .11110 &   .01239 &      .00137 \\
48 &   .13625 &  .01858 &   .00245 &     185 &  .11069 &   .01230 &      .00136 \\
50 &   .13529 &  .01832 &   .00241 &     190 &  .11029 &   .01221 &      .00134 \\
52 &   .13438 &  .01808 &   .00236 &     195 &  .10991 &   .01212 &      .00133 \\
54 &   .13351 &  .01785 &   .00232 &     200 &  .10954 &   .01204 &      .00132 \\
\hline
\end{tabular}
\end{center}
\end{table}

\begin{table}[h]
\caption{The three-loop Pade approximated global  analytic coupling ${\agoth}_{1}(s)$ and
the corresponding
global ``analyticized  powers'' ${\agoth}_2(s)$ and ${\agoth}_3(s)$
       as a functions of the energy $\sqrt s$ for $\Lambda_{3}=400$
$MeV$. }
\vspace{0.2cm}
\begin{center}
\begin{tabular}{|l|l|l|l|l|l|l|l|}
\hline ${\sqrt s}$ $GeV$ &   ${\agoth}_{1}(s)$ & $ {\agoth}_{2}(s)$ &
${\agoth}_3(s)$ & ${\sqrt s }$ $GeV$ &  ${\agoth }_{1}(s)$ & $ {\agoth}_{2}(s)$& ${\agoth}_3(s)$ \\
\hline
1 &  .36427 &    .10048 &  .02061  &         58&   .12898 &   .01628 & .00201  \\
2 &  .28168 &    .06873 &  .01440  &         60&   .12828 &    .01611&.00198    \\
3 &  .24639 &    .05489 &  .01100  &         62&   .12761 &    .01595& .00195   \\
4 &  .22592 &    .04723 &  .00911  &         64&   .12697 &    .01579&  .00192  \\
5 &  .21256 &    .04228 &  .00785  &         66&   .12635 &    .01564&  .00189  \\
6 &  .20288 &    .03876 &  .00696  &         68&   .12576 &    .01550&  .00187  \\
7 &  .19533 &    .03610 &  .00630  &         70&   .12519 &    .01537&   .00184 \\
8 &  .18922 &    .03400 &  .00579  &         72&   .12464 &    .01523&  .00182   \\
9 &  .18413 &    .03230 &  .00539  &         74&   .12411 &    .01511&  .00180   \\
10&   .17980&     .03087&   .00505 &         76&   .12360 &    .01499&  .00178   \\
11&   .17605&     .02966&   .00477 &         78&   .12311 &    .01487&  .00176   \\
12 &  .17276 &    .02861 &  .00453 &         80 &  .12263 &    .01476 & .00174   \\
13 &  .16984 &    .02769 &  .00433 &         82 &  .12217 &    .01465 & .00172   \\
14 &  .16722 &    .02688 &  .00415 &         84 &  .12172 &    .01455 & .00170    \\
15 &  .16485 &    .02616 &  .00399 &         86 &  .12129 &    .01445 & .00168    \\
16 &  .16269 &    .02551 &  .00385 &         88 &  .12087 &    .01435 & .00167    \\
17 &  .16072 &    .02492 &  .00372 &         90 &  .12046 &    .01425 & .00165    \\
18 &  .15890 &    .02438 &  .00360 &         92 &  .12006 &    .01416 & .00164    \\
19 &  .15721 &    .02388 &  .00350 &         94 &  .11968 &    .01407 & .00162    \\
20 &  .15565 &    .02343 &  .00340 &         96 &  .11930 &    .01399 & .00161     \\
21 &  .15419 &    .02301 &  .00332 &         98 &  .11894 &    .01391 & .00159     \\
22 &  .15282 &    .02262 &  .00323 &         100&   .1185 &     .01382&   .00158   \\
23 &  .15154 &    .02225 &  .00316 &         105 &  .11773  &   .01363 &  .00155   \\
24 &  .15033 &    .02191 &  .00309 &         110 &  .11694  &   .01345 &  .00152   \\
25 &  .14919 &    .02159 &  .00302 &         115 &  .11618  &   .01328 &  .00149     \\
26 &  .14811 &    .02129 &  .00296 &         120 &  .11547  &   .01313 &  .00146      \\
27 &  .14708 &    .02101 &  .00291 &         125 &  .11480  &   .01298 &  .00144      \\
28 &  .14610 &    .02074 &  .00285 &         130 &  .11415  &   .01283 &  .00142      \\
29 &  .14518 &    .02048 &  .00280 &         135 &  .11354  &   .01270 &  .00139      \\
30 &  .14429 &    .02024 &  .00275 &         140 &  .11296  &   .01257 &  .00138      \\
32 &  .14263 &   .01979  &   .00267&         145 &  .11241  &   .01245 &  .00136      \\
34 &  .14111 &   .01939  &   .00259&          150&   .11188 &   .01234 &  .00134       \\
36 &  .13970 &   .01902  &   .00252&       155   &   .11137 &   .01223 &  .00132       \\
38 &  .13840 &   .01867  &   .00245  &      160 &  .11088  &    .01212 &  .00130       \\
40 &  .13718 &   .01836  &   .00239  &      165 &  .11041 &     .01202 &  .00129       \\
42 &  .13605 &   .01806  &  .00233   &      170 &  .10996  &    .01193 &  .00127       \\
44 &  .13498 &   .01779  &   .00228  &       175&   .10956 &     .01184&   .00126      \\
46 &  .13398 &   .01753  &  .00223   &      180 &  .10918  &    .01176 &  .00125        \\
48 &  .13303 &   .01729  &  .00219   &      185 &  .10881  &    .01168 &  .00123        \\
50 &  .13213 &   .01707  &  .00215   &      190 &  .10845  &    .01161 &  .00122        \\
52 &  .13129 &   .01685  &  .00211   &      195 &  .10811  &    .01154 &  .00121        \\
54 &  .13048 &   .01665  &  .00207   &      200 &  .10777  &    .01147 &  .00120        \\
\hline
\end{tabular}
\end{center}
\end{table}

\begin{table}[h]
\caption{The three-loop Pade approximated global  analytic
coupling ${\cal A}_{1}(Q^2)$ and the corresponding global
``analyticized  powers'' ${\cal A}_2(Q^2)$ and ${\cal A}_3(Q^2)$
         as a functions of the momentum transfer Q for $\Lambda_{3}=450$ $MeV$.
         The matching conditions give:
         $\Lambda_{4}=411.1$ $MeV$, $\Lambda_{5}=307.5$ $MeV$ and
$\Lambda_{6}=135.4$ MeV.} \vspace{0.2cm}
\begin{center}
\begin{tabular}{|l|l|l|l|l|l|l|l|}
\hline $Q$ $GeV$&    ${\cal A}_{1}(Q^2)$&  $ {\cal A}_2(Q^2)$ &
   ${\cal A}_3(Q^2)$& $Q$ $GeV$ & ${\cal A}_{1}(Q^2)$&
    ${\cal A}_2(Q^2)$&   ${\cal A}_3(Q^2)$ \\
\hline
1   &  .40668 &    .10611 &   .01692 &  38 &  .146322  &   .02133  &      .00306   \\
2   &  .31810 &    .07993 &   .01488 &  40 &  .144927  &   .02093  &      .00298    \\
3   &  .27716 &    .06602 &   .01264 &  42 &  .143626  &   .02057  &      .00291    \\
4   &  .25295 &    .05739 &   .01098 &  44 &  .142408  &   .02022  &      .00284    \\
5   &  .23638 &    .05146 &   .00975 &  46 &  .141263  &   .01991  &      .00277     \\
6   &  .22419 &    .04710 &   .00880 &  48 &   .140186 &    .01961 &       .00271    \\
7   &  .21459 &   .04374  &  .00806  &  50 &    .139168&     .01933&        .00266   \\
8   &  .20715 &   .04106  &  .00746  &  52 &    .138205&     .01907&        .00260   \\
9   &  .20086 &   .03886  &  .00696  &   54&   .137291 &    .01882 &       .00255    \\
10  & .19554  &  .03702   &   .00654 &   56&   .136422 &    .01858 &       .00251    \\
11  & .19095  &  .03545   &    .00618&   58&   .135595 &    .01836 &       .00246    \\
12  & .18695  &  .03410   &    .00587&   60&   .134806 &    .01815 &       .00243  \\
13  & .18341  &  .03291   &    .00560&   62&   .134052 &    .01795 &       .00238    \\
14  & .18024  &  .03185   &    .00537&   64&   .133330 &    .01776 &       .00235    \\
15  & .17740  &  .03092   &    .00515&   66&   .132639 &    .01758 &       .00231     \\
16  & .17481  &  .03007   &    .00496&   68&   .131975 &    .01740 &        .00228    \\
17  & .17245  &  .02931   &    .00479&   70&   .131338 &    .01724 &        .00225    \\
18  & .17029  &  .02861   &    .00464&   72&   .130725 &    .01708 &        .00222     \\
19  & .16829  &  .02797   &  .00450  &   74&   .130134 &    .01693 &        .00219     \\
20  & .16643  &  .02739   &  .00437  &   76&   .129565 &    .01678 &        .00216     \\
21  & .16471  &  .02685   &  .00425  &   78&    .12885 &     .01664&        .00213     \\
22  & .16310  &  .02634   &  .00414  &   80&   .128485 &     .01650&        .00211     \\
23  & .16159  &  .02587   &  .00403  &   82&   .127972 &     .01637&        .00208     \\
24  & .16017  &  .02544   &   .00394 &   84&   .127476 &     .01625&        .00206     \\
25  & .15883  &  .02503   &   .00385 &   86&   .126995 &     .01613&        .00204     \\
26  & .15756  &  .02464   &   .00377 &   88&   .126530 &     .01601&        .00201     \\
27  & .15637  &  .02428   &   .00369 &   90&   .126078 &     .01590&        .00199     \\
28  & .15523  &  .02394   &   .00361 &   92&   .125640 &     .01579&        .00197      \\
29  & .15415  &  .02362   &   .00354 &   94&   .125215 &     .01568&        .00195      \\
30  & .15312  &  .02331   &   .00348 &   96&   .124801 &     .01558&        .00193      \\
32  & .15120  &   .02274  &    .00336&   98&   .124238 &     .01548&        .00192       \\
34  & .14944  &   .02223  &    .00325&   100&  .12384  &    .01538 &       .00190        \\
36  & .14782  &   .02176  &    .00315&     &           &           &                    \\
 \hline
 \end{tabular}
 \end{center}
 \end{table}

\begin{table}[h]
\caption{The three-loop Pade approximated global  analytic coupling ${\agoth}_{1}(s)$ and the corresponding
``analyticized  powers'' ${\agoth}_2(s)$ and ${\agoth}_3(s)$
       as a functions of the energy $\sqrt s $ for $\Lambda_{3}=450$
         $MeV$.}

\vspace{0.2cm}
\begin{center}
\begin{tabular}{|l|l|l|l|l|l|l|l|}
\hline ${\sqrt s}$ $GeV$ &   ${\agoth}_{1}(s)$ & $ {\agoth}_{2}(s)$ &
${\agoth}_3(s)$ & ${\sqrt s }$ $GeV$ &  ${\agoth }_{1}(s)$ & $ {\agoth}_{2}(s)$& ${\agoth}_3(s)$
\\
 \hline
1  &  .38512  & .10857  & .02191 &         58&   .13226 &  .01710 &  .00216  \\
2  &  .29624  & .07474  & .01590 &         60&   .13152 &  .01692 &  .00212  \\
3  &  .25798  & .05953  & .01221 &         62&   .13082 &  .01674 &  .00209  \\
4  &  .23585  & .05108  & .01012 &         64&   .13014 &  .01657 &  .00206  \\
5  &  .22142  & .04560  & .00871 &         66&   .12950 &  .01641 &  .00203  \\
6  &  .21096  & .04169  & .00770 &         68&   .12887 &  .01626 &  .00200   \\
7  &  .20283  & .03875  & .00696 &         70&   .12828 &  .01612 &  .00198  \\
8  &  .19627  & .03643  & .00638 &         72&   .12770 &  .01598 &  .00195   \\
9  &  .19081  & .03455  & .00592 &         74&   .12715 &  .01584 &  .00193   \\
10 &  .18617  & .03298  & .00555 &        76 &  .12661  & .01571  & .00191    \\
11 &  .18216  & .03165  & .00523 &        78 &  .12609  & .01559  & .00188   \\
12 &  .17865  & .03050  & .00497 &        80 &  .12559  & .01547  & .00186   \\
13 &  .17553  & .02949  & .00473 &        82 &  .12511  & .01535  & .00184   \\
14 &  .17273  & .02861  & .00453 &        84 &  .12464  & .01524  & .00182   \\
15 &  .17021  & .02781  & .00436 &        86 &  .12419  & .01513  & .00180   \\
16 &  .16792  & .02710  & .00420 &        88 &  .12374  & .01503  & .00179    \\
17 &  .16581  & .02646  & .00405 &        90 &  .12332  & .01493  & .00177   \\
18 &  .16388  & .02587  & .00393 &        92 &  .12290  & .01483  & .00175    \\
19 &  .16209  & .02533  & .00381 &        94 &  .12250  & .01473  & .00174    \\
20 &  .16043  & .02483  & .00370 &        96 &  .12211  & .01464  & .00172    \\
21 &  .15888  & .02438  & .00360 &        98 &  .12172  & .01455  & .00170       \\
22 &  .15743  & .02395  & .00351 &        100&   .12135 & .01447  & .00169       \\
23 &  .15607  & .02355  & .00343 &        105&   .12046 &  .01426 &  .00165      \\
24 &  .15479  & .02318  & .00335 &        110&   .11963 &  .01407 &  .00162      \\
25 &  .15358  & .02284  & .00328 &        115&   .11884 &  .01389 &  .00159      \\
26 &  .15243  & .02251  & .00321 &        120&   .11809 &  .01372 &  .00156       \\
27 &  .15135  & .02220  & .00315 &        125&   .11739 &  .01356 &  .00154      \\
28 &  .15032  & .02191  & .00309 &        130&   .11672 &  .01341 &  .00151       \\
29 &  .14933  & .02163  & .00303 &        135&   .11608 &  .01327 &  .00149       \\
30 &  .14840  & .02137  & .00298 &        140&   .11547 &  .01313 &  .00147       \\
32 &  .14665  & .02089  & .00288 &        145&   .11489 &  .01300 &  .00145      \\
34 &  .14504  & .02045  & .00280 &        150&   .11434 &  .01288 &  .00143      \\
36 &  .14355  & .02005  & .00272 &        155&   .11381 &  .01276 &  .00141      \\
38 &  .14218  & .01968  & .00264 &        160&   .11330 &  .01265 &  .00139      \\
40 &  .14089  & .01933  & .00258 &        165&   .11281 &  .01255 &  .00137      \\
42 &  .13970  & .01902  & .00252 &        170&   .11233 &  .01244 &  .00136       \\
44 &  .13857  & .01872  & .00246 &        175&   .11192 &  .01235 &  .00134      \\
46 &  .13752  & .01845  & .00241 &        180&   .11152 &  .01227 &  .00133       \\
48 &  .13652  & .01819  & .00236 &        185&   .11114 &  .01218 &  .00131       \\
50 &  .13558  & .01794  & .00231 &        190&   .11077 &  .01210 &  .00130       \\
52 &   .13468 &  .01772 &  .00227&        195 &  .11041 &  .01202 &  .00129         \\
54 &  .13383  & .01750  & .00223 &        200&   .11006 &  .01195 &  .00128        \\
\hline
\end{tabular}
\end{center}
\end{table}
\end{document}